\documentclass[prd,aps,superscriptaddress,twocolumn,floatfix,nofootinbib]{revtex4-1}
\pdfoutput=1

\usepackage{amsfonts}
\usepackage{amsmath}
\usepackage{amssymb}
\usepackage{bm}
\usepackage{dcolumn}
\usepackage{graphicx}   
\usepackage[latin1]{inputenc}
\usepackage{latexsym}
\usepackage{rotating}
\usepackage{hyperref}
\usepackage{graphicx}
\usepackage{color}

\newcommand\be{\begin{equation}}
\newcommand\ba{\begin{eqnarray}}
\newcommand\ee{\end{equation}}
\newcommand\ea{\end{eqnarray}}

\begin{document}

\title{A Non-Inflationary Axion and ALP Misalignment Mechanism}

\author{Robert Brandenberger}
\email{rhb@physics.mcgill.ca}
\affiliation{Department of Physics, McGill University, Montr\'{e}al,
  QC, H3A 2T8, Canada}

%%%%%%%%%%%%%%%%%%%%%%%%%%%%%%%%%%%%%%%%%%

\begin{abstract}

Based on considerations of quantum gravity, global symmetries which lead to axions as pseudo Nambu-Goldstone bosons after low scale symmetry breaking cannot be exact at the Planck scale. Here, we show that Planck-suppressed terms which yield this symmetry breaking may provide a non-inflationary misalignment mechanism which can generate coherent oscillations of the axion and axion-like particle (ALP) fields at low temperatures.

\end{abstract}

\maketitle

%%%%%%%%%%%%%%%%%%%%%%%%%%%%%%%%%%%%%%%%%%
\section{Introduction} 
\label{sec:intro}

Axions and axion-like pseudo-scalar particles (ALPs) are playing an increasing role in particle physics and cosmology (see e.g. \cite{axionrev} for reviews).  The QCD axion \cite{axion, axion2} has been postulated as a solution of the strong CP problem and can yield a candidate for particle dark matter \cite{axionDM}.  Axions arise generically in low energy effective field theories coming from string theory \cite{Witten}. ALPs have been put forwards as candidates for dark matter and dark energy (see e.g. \cite{ALPrev} for reviews). It has recently been pointed out \cite{Bgen} that ALP fields which are coherently oscillating over cosmological scales at the time of recombination can generate primordial magnetic field on the scale of voids of sufficient magnitude to exceed the observational upper bounds (see e.g. \cite{Durrer} for a review of the observational status), and that ALP fields coherently oscillating on galactic scales can yield the required amplitude of Lyman-Werner radiation to enable \cite{JH1} formation of super-massive black holes at high redshifts via the ``Direct Collapse Black Hole'' formation scenario, in particular if there is a scaling distribution of cosmic string loops which provides a sufficient number of nonlinear seed fluctuations \cite{JH2}.
 
Axions and ALPs result from the spontaneous breaking of a global $U(1)$ symmetry \cite{PQ}.  For example, the QCD axion is the angular variable $\theta$ of a complex scalar field $\phi$ with a potential $V(|\phi|)$ which has a nontrivial minimum at $|\phi| = f_a$.  The energy scale of the spontaneous breaking of the $U(1)$ symmetry is called the Peccei-Quinn symmetry breaking scale.  By causality \cite{Kibble}, the angle $\theta$ cannot be correlated on super-Hubble scales. Non-perturbative instanton effects break the $U(1)$ rotational symmetry of the potential and induce a potential $\delta V$ and mass for the angular component.  
 
Axions and ALPs are often assumed to be coherently oscillating on cosmological scales at late times.  These oscillations are generated from the assumption that before late time symmetry breaking the angular variable $\theta$ is displaced from the minimum of the late time potential $\delta V$ (which we can take to be $\theta = 0$ without loss of generality) uniformly in space over cosmological scales.

What is the origin of this {\it misalignment}? The usual answer given is that the misalignment is a result of the evolution in an early phase of inflation, assuming that the Peccei-Quinn symmetry breaking occurs before or during inflation \footnote{It the symmetry is broken only after inflation, then causality dictates that the angle $\theta$ must be uncorrelated on scales larger than the horizon at the time of the phase transition.} Because of quantum fluctuations,  at the beginning of inflation $\theta$ has a random distribution with a characteristic coherence scale given by the Hubble radius.  This coherence scale of $\theta$ increases exponentially during inflation, effectively producing a value of $\theta$ which is almost uniform and has a value of order one relative to the minimum of the late time potential over the current cosmological horizon.  

However, cosmological inflation is only one of several possible early universe scenarios. Alternatives include bouncing models (see e.g. \cite{Patrick} for a review), in particular the Ekpyrotic scenario \cite{Ekp}, and emergent models such as String Gas Cosmology \cite{BV} and matrix cosmology \cite{matrix} (see e.g. \cite{RBrev} for a review of early universe scenarios).  Recently, a number of conceptual challenges have been raised which indicate the standard slow-roll inflation cannot emerge at the level of an effective field theory. These include ``swampland'' arguments \cite{dS} which show that standard slow-roll inflation cannot emerge from the corners of string theory tractable in terms of effective field theories,  and the ``Trans-Planckian Censorship Criterion'' \cite{TCC1, TCC2} which indicates that a long period of inflation leads to unitarity problems \cite{TCC3} and possible violations of the second law of thermodynamics \cite{BB}. 

In light of these problems the question arises as to whether there is a different mechanism which can generate the misalignment of $\theta$ over cosmological scales.
In this Letter we shall present a new mechanism to produce this misalignment which is based on the fact that the global symmetries which yield axions and ALPs must be broken at the Planck scale (see e.g. \cite{noglobal} for an early review, \cite{swamprev} for reviews in the context of string theory, and \cite{noglobal-axion} for original articles related to the axion).  The basic idea is very simple: the global symmetry violating terms set $\theta$ to a particular value in the very early universe when the global symmetry violating terms dominate the action.  There is no reason that the value of $\theta$ associated with this value of $\phi$ which we denote by $\theta_i$ is the same as the value of $\theta$ (which we take to be $\theta = 0$ which is singled out by the late time instantons which violate the global symmetry at late times. Due to Hubble friction, the value of $\theta$ is frozen to $\theta = \theta_i$ until the Hubble constant drops below the mass $m_a$ of the ALP at late times. At that time, oscillations of $\theta$ about $\theta = 0$ can start, and these oscillations are coherent over all of space.

%%%%%%%%%%%%%%%%%%%%%%%%%%%%%%%%%%%%%%%%%%%
\section{Analysis}

We will assume that the ALP field is the phase of a complex scalar field $\phi$ in the same way that the QCD axion is the phase of the Peccei-Quinn field.  The global $U(1)$ symmetry is
\be
\phi \, \rightarrow \, \phi e^{i\alpha} \, ,
\ee
where $\alpha$ is an arbitrary phase.  Taking into account both the instanton effects which break the symmetry and generate a minimum of the phase $\theta$ of $\phi$ at low energies, and the terms which yield explicit global symmetry violation at the Planck scale and which are Planck-suppressed at low energies, the total potential for $\phi$ can be written as
\be \label{full}
V(\phi) \, = \, V_0(\phi) + \delta V_{l}(\phi) + V_p(\phi) \, ,
\ee
where
\be \label{dominant}
V_0(\phi) \, = \, \lambda (|\phi|^2 - f_a^2)^2
\ee
is the potential which yields spontaneous symmetry breaking at temperatures lower than a scale (the generalization of the Peccei-Quinn (PQ) symmetry breaking scale) which is of the order $f_a$, 
\be \label{instanton}
\delta V_{l}(\theta) \, = \, \frac{1}{2} m_a^2 f_a^2 \theta^2 + {\cal{O}} (\theta^4)
\ee
is the global symmetry breaking potential generated by instantons which generates the ALP mass at low energies, and $V_p(\phi)$ denotes the terms in the potential which yield global symmetry violation at high scales, and which are Planck suppressed at low energies. Note that we have written $\delta V_l$ as a function of the angle only, assuming that the radial piece of $\phi$ is frozen at the value $f_a$ (which will be the case below the PQ symmetry breaking scale). This potential is periodic in $\theta$. In order to avoid an axion domain wall problem, (see e.g. \cite{Sarkar}) we choose the period to be $2\pi$. We have expanded this potential about the value $\theta = 0$ which we assume to be the minimum of the potential.  

In light of the global symmetry of (\ref{dominant}) we can write
\be
\phi \, = \phi_r e^{i\theta} \,  ,
\ee
where $\phi_r$ is the radial field.  Expanding (\ref{dominant}) of the potential about $\phi_r = f_a$ we find that the mass of the radial component of $\phi$ is
\be
m_r \, = \, 2 \sqrt{2} \sqrt{\lambda} f_a \, .
\ee
At energies larger than $m_r$,  we expect that the global symmetry is restored, i.e. that on average $\phi_r = 0$. On the other hand, at energies much smaller than $m_r$, we expect $\phi_r$ to be frozen at the value $\phi_r = f_a$.  For the QCD axion, the energy $m_r$ corresponds to the Peccei-Quinn symmetry breaking scale.

Planck scale effects must break the global $U(1)$ symmetry. We suggest two parametrizations of this effect. The first is via explicity symmetry breaking terms in the ALP sector
\be \label{same}
V(\phi_r, \theta) \, = \, \frac{\phi_r^{4 + 2n}}{m_{pl}^{2n}} \bigl[1 - {\rm{cos}}(\theta + \beta) \bigr] \, ,
\ee
where $n$ is an integer, $m_{pl}$ is the Planck mass, and $\beta$ is a phase. There is no reason why $\beta$ should vanish (i.e.  that the distinguished angle of the Planck suppressed terms should correspond to the angle of the minumum of the instanton terms $\delta V$).  In the case of the axion, the integer $n$ must be greater or equal to $n = 6$ in order for the axion to be able to solve the strong CP problem \cite{noglobal-axion} (this is the ``axion quality problem'' - see e.g. \cite{Dine} for a recent review).  For ALPs this issue is not present and thus we can consider the simplest case, namely $n = 1$.

It is also possible that the global symmetry is broken in a different sector and mediated to the ALP sector via thermal effects. In this case, we can consider a potential of the form ($T$ is the temperature)
\be \label{hidden}
V_T(T, \theta) \, = \, \frac{T^{2 + 2n}}{m_{pl}^{2n}} f_a^2 \bigl[1 - {\rm{cos}}(\theta + \beta) \bigr] \, ,
\ee
when restricted to values of $\phi$ with $|\phi| = f_a$.  To justify the above effective potential,  assume that $\phi$ couples to a hidden sector field $\chi$ via an interaction term
\be
V_{int} \, = \, \frac{\chi^{2 + 2n}}{m_{pl}^{2n}} |\phi|^2 \bigl[1 - {\rm{cos}}(\theta + \beta) \bigr]
\ee
such that the total Lagrangian violates the global symmetry.  Without much loss of generality we can take $\chi$ to have a self interaction potential of the form 
\be
V_{\chi} \, = \, \lambda_{\chi} \chi^4
\ee
with $\lambda_{\chi} \sim 1$.  Then , if$\chi$ is in thermal equilibrium we have $\chi^4 \sim T^4$, and integrating out $\chi$ leads to an effective potential for $\phi$ for the form (\ref{hidden}) \footnote{A similar argument has been used to obtain the finite temperature correction to the effective potential of a scalar field, see e.g. \cite{RBfT}.}
  
 In the case of the parametrization (\ref{same}), the Planck scale global symmetry breaking terms produce a mass for the axion $\theta$. Unless the offset angle $\beta$ is chosen to be extremely small, the minimum of the axion potential is $\theta = - \beta$ independent of the energy of the background cosmology.  This does not lead to any misaligment. In fact, in the case of the QCD axion, these symmetry breaking terms would destroy to ability of the axion to solve the strong CP problem. 
 
To obtain a viable cosmological misalignment, we will hence focus on the symmetry breaking potential of the form (\ref{hidden}).  At high temperatures $m_r > T \gg m_a$,  the thermal effects localize $\theta$ at the value
\be
\theta \, = \, - \beta \, ,
\ee
as can be immediately seen from (\ref{hidden}).  As the temperature drops, the value of $\theta$ which minimizes the total potential
\be
V_{total}(T, \theta) \, = \, \frac{T^{2 + 2n}}{m_{pl}^{2n}} f_a^2 \bigl[1 - {\rm{cos}}(\theta + \beta) \bigr] + \frac{1}{2} m_a^2 \theta^2 
\ee
continuously moves from $\theta = - \beta$ to $\theta = 0$. However, due to Hubble friction $\theta$ is frozen until the time when
\be
H(T) \, = \, m_a \, ,
\ee
when $\theta$ will begin oscillations about $\theta = 0$ coherently over space (modulo thermal fluctuations). 

It is easy to argue that any coherent thermal fluctuations are suppressed on large length scales $R$ as
\be
\delta \theta \, \sim \, R^{-3/2} \, .
\ee
Specifically, considering a fluctuation region of radius $R$ to have thermal energy $T$, we find that the thermal fluctuations are suppressed on scales
\be \label{cond}
R \, \gg \, \bigl( \frac{m_{pl}}{T} \bigr)^{2n/3} T^{-1/3} f_a^{-2/3} \, .
\ee
To gain a feeling of what this scale is, consider the temperature when the offset of $\theta$ is frozen in to be $10^{-m} m_{pl}$ and $f_a = m_{pl}$. In this case (\ref{cond}) becomes
\be
R \, \gg \, 10^{\frac{2}{3} m n - m/3} m_{pl}^{-1} \, ,
\ee
which is microscopical.
  
\section{Discussion}  

 Axions and ALPs often arise from the breaking of a global symmetry.  Global symmetries are, however, thought to be broken by quantum gravity effects. We have here proposed a mechanism by which this global symmetry breaking at the Planck scale can lead to late time coherent oscillations of the axion and ALP fields. This misalignment mechanism does not require the early universe to have undergone a period of inflation, and in light of the conceptual problems of standard inflation  this is an important advantage.
 
 The proposed mechanism for generating coherent late time oscillations of an axion or ALP field is not independent of the form of the global symmetry breaking, but a generic effective potential of the form (\ref{hidden}) which results from global symmetry breaking via interactions of $\phi$ with a hidden sector field $\chi$ will work.
\\
%%%%%%%%%%%%%%%%%%%%%%%%%%%%%%%%%%%%%%%%%%
\begin{acknowledgements}

I wish to thank Simon Caron-Huot and Davide Racco for useful discussions,  and Jiao Hao, Juerg Froehlich and Vahid Kamali for collaboration on related topics. My research is supported in part by an NSERC Discovery Grant and by the Canada Research Chair program.    

\end{acknowledgements}
%%%%%%%%%%%%%%%%%%%%%%%%%%%%%%%%%%%%%%%%%

%%%%%%%%%%%%%%%%%%%%%%%%%%%%%%%%%%%%%%%%%%


\begin{thebibliography}{99}

\bibitem{axionrev}
%\cite{Kim:1986ax}
J.~E.~Kim,
``Light Pseudoscalars, Particle Physics and Cosmology,''
Phys. Rept. \textbf{150}, 1-177 (1987)
doi:10.1016/0370-1573(87)90017-2;\\
%\cite{Marsh:2015xka}
D.~J.~E.~Marsh,
``Axion Cosmology,''
Phys. Rept. \textbf{643}, 1-79 (2016)
doi:10.1016/j.physrep.2016.06.005
[arXiv:1510.07633 [astro-ph.CO]].

\bibitem{axion}
%\cite{Weinberg:1977ma}
S.~Weinberg,
``A New Light Boson?,''
Phys. Rev. Lett. \textbf{40}, 223-226 (1978)
doi:10.1103/PhysRevLett.40.223;\\
%\cite{Wilczek:1977pj}
F.~Wilczek,
``Problem of Strong  $P$  and  $T$  Invariance in the Presence of Instantons,''
Phys. Rev. Lett. \textbf{40}, 279-282 (1978)
doi:10.1103/PhysRevLett.40.279

\bibitem{axion2}
%\cite{Dine:1981rt}
M.~Dine, W.~Fischler and M.~Srednicki,
``A Simple Solution to the Strong CP Problem with a Harmless Axion,''
Phys. Lett. B \textbf{104}, 199-202 (1981)
doi:10.1016/0370-2693(81)90590-6;\\
%\cite{Zhitnitsky:1980tq}
A.~R.~Zhitnitsky,
``On Possible Suppression of the Axion Hadron Interactions. (In Russian),''
Sov. J. Nucl. Phys. \textbf{31}, 260 (1980);\\
%\cite{Dine:1982ah}
M.~Dine and W.~Fischler,
``The Not So Harmless Axion,''
Phys. Lett. B \textbf{120}, 137-141 (1983)
doi:10.1016/0370-2693(83)90639-1.

\bibitem{axionDM}
%\cite{Preskill:1982cy}
J.~Preskill, M.~B.~Wise and F.~Wilczek,
``Cosmology of the Invisible Axion,''
Phys. Lett. B \textbf{120}, 127-132 (1983)
doi:10.1016/0370-2693(83)90637-8

\bibitem{Witten}
%\cite{Svrcek:2006yi}
P.~Svrcek and E.~Witten,
``Axions In String Theory,''
JHEP \textbf{06}, 051 (2006)
doi:10.1088/1126-6708/2006/06/051
[arXiv:hep-th/0605206 [hep-th]].

\bibitem{ALPrev}
%\cite{Ferreira:2020fam}
E.~G.~M.~Ferreira,
``Ultra-light dark matter,''
Astron. Astrophys. Rev. \textbf{29}, no.1, 7 (2021)
doi:10.1007/s00159-021-00135-6
[arXiv:2005.03254 [astro-ph.CO]];\\
%\cite{Hui:2016ltb}
L.~Hui, J.~P.~Ostriker, S.~Tremaine and E.~Witten,
``Ultralight scalars as cosmological dark matter,''
Phys. Rev. D \textbf{95}, no.4, 043541 (2017)
doi:10.1103/PhysRevD.95.043541
[arXiv:1610.08297 [astro-ph.CO]].

\bibitem{Bgen}
%\cite{Joyce:1997uy}
R.~Brandenberger, J.~Fr\"ohlich and H.~Jiao,
``Cosmological Magnetic Fields from Ultralight Dark Matter,''
[arXiv:2502.19310 [hep-ph]].

\bibitem{Durrer}
%\cite{Durrer:2013pga}
R.~Durrer and A.~Neronov,
``Cosmological Magnetic Fields: Their Generation, Evolution and Observation,''
Astron. Astrophys. Rev. \textbf{21}, 62 (2013)
doi:10.1007/s00159-013-0062-7
[arXiv:1303.7121 [astro-ph.CO]].

\bibitem{JH1}
%\cite{Jiao:2025kpn}
H.~Jiao, R.~Brandenberger and V.~Kamali,
``Direct Collapse Supermassive Black Holes from Ultralight Dark Matter,''
[arXiv:2503.19414 [astro-ph.CO]].

\bibitem{JH2}
%\cite{Cyr:2022urs}
B.~Cyr, H.~Jiao and R.~Brandenberger,
``Massive black holes at high redshifts from superconducting cosmic strings,''
Mon. Not. Roy. Astron. Soc. \textbf{517}, no.2, 2221-2230 (2022)
doi:10.1093/mnras/stac1939
[arXiv:2202.01799 [astro-ph.CO]];\\
%\cite{Jiao:2023wcn}
H.~Jiao, R.~Brandenberger and A.~Refregier,
``Early structure formation from cosmic string loops in light of early JWST observations,''
Phys. Rev. D \textbf{108} (2023) no.4, 043510
doi:10.1103/PhysRevD.108.043510
[arXiv:2304.06429 [astro-ph.CO]];\\
%\cite{Jiao:2024rcr}
H.~Jiao, R.~Brandenberger and A.~Refregier,
``N-body simulation of early structure formation from cosmic string loops,''
Phys. Rev. D \textbf{109}, no.12, 123524 (2024)
doi:10.1103/PhysRevD.109.123524
[arXiv:2402.06235 [astro-ph.CO]].

\bibitem{PQ}
%\cite{Peccei:1977hh}
R.~D.~Peccei and H.~R.~Quinn,
``CP Conservation in the Presence of Instantons,''
Phys. Rev. Lett. \textbf{38}, 1440-1443 (1977)
doi:10.1103/PhysRevLett.38.1440

\bibitem{Kibble}
%\cite{Kibble:1980mv}
T.~W.~B.~Kibble,
``Some Implications of a Cosmological Phase Transition,''
Phys. Rept. \textbf{67}, 183 (1980)
doi:10.1016/0370-1573(80)90091-5

\bibitem{Patrick}
R.~Brandenberger and P.~Peter,
 ``Bouncing Cosmologies: Progress and Problems,''
 Found.\ Phys.\  {\bf 47}, no. 6, 797 (2017)
% doi:10.1007/s10701-016-0057-0
 [arXiv:1603.05834 [hep-th]].
 %%CITATION = doi:10.1007/s10701-016-0057-0;%%
 
\bibitem{Ekp}
  J.~Khoury, B.~A.~Ovrut, P.~J.~Steinhardt and N.~Turok,
 ``The Ekpyrotic universe: Colliding branes and the origin of the hot big
 bang,''
 Phys.\ Rev.\ D {\bf 64}, 123522 (2001) [hep-th/0103239].
 %%CITATION = HEP-TH/0103239;%%

\bibitem{BV}
 R.~H.~Brandenberger and C.~Vafa,
	``Superstrings In The Early Universe,'' 
	Nucl.\ Phys.\ B {\bf 316}, 391 (1989).
	%%CITATION = NUPHA,B316,391;%%
	 
\bibitem{matrix}
%\cite{Brahma:2021tkh}
S.~Brahma, R.~Brandenberger and S.~Laliberte,
``Emergent cosmology from matrix theory,''
JHEP \textbf{03}, 067 (2022)
[arXiv:2107.11512 [hep-th]];\\
%\cite{Brahma:2022dsd}
S.~Brahma, R.~Brandenberger and S.~Laliberte,
``Emergent metric space-time from matrix theory,''
JHEP \textbf{09}, 031 (2022)
doi:10.1007/JHEP09(2022)031
[arXiv:2206.12468 [hep-th]];\\
  %\cite{Brahma:2022hjv}
S.~Brahma, R.~Brandenberger and S.~Laliberte,
``Emergent early universe cosmology from BFSS matrix theory,''
Int. J. Mod. Phys. D \textbf{31}, no.14, 2242004 (2022)
doi:10.1142/S0218271822420044
[arXiv:2205.06016 [hep-th]];\\
%\cite{Brahma:2022ikl}
S.~Brahma, R.~Brandenberger and S.~Laliberte,
``BFSS Matrix Model Cosmology: Progress and Challenges,''
[arXiv:2210.07288 [hep-th]].

\bibitem{RBrev}
%\cite{Brandenberger:2010bpq}
R.~H.~Brandenberger,
``Introduction to Early Universe Cosmology,''
PoS \textbf{ICFI2010}, 001 (2010)
doi:10.22323/1.124.0001
[arXiv:1103.2271 [astro-ph.CO]];\\
%\cite{Brandenberger:2023ver}
R.~Brandenberger,
``Superstring cosmology \textemdash{} a complementary review,''
JCAP \textbf{11}, 019 (2023)
doi:10.1088/1475-7516/2023/11/019
[arXiv:2306.12458 [hep-th]].

\bibitem{dS}
 G.~Obied, H.~Ooguri, L.~Spodyneiko and C.~Vafa,
 ``De Sitter Space and the Swampland,''
 arXiv:1806.08362 [hep-th].
 %%CITATION = ARXIV:1806.08362;%%

\bibitem{TCC1}
 A.~Bedroya and C.~Vafa,
 ``Trans-Planckian Censorship and the Swampland,''
JHEP {\bf 2009}, 123 (2020)
  [arXiv:1909.11063 [hep-th]].
  %%CITATION = doi:10.1007/JHEP09(2020)123;%%

\bibitem{TCC2}
 A.~Bedroya, R.~Brandenberger, M.~Loverde and C.~Vafa,
 ``Trans-Planckian Censorship and Inflationary Cosmology,''
 Phys.\ Rev.\ D {\bf 101}, no. 10, 103502 (2020)
 [arXiv:1909.11106 [hep-th]].
 %%CITATION = doi:10.1103/PhysRevD.101.103502;%%
 
\bibitem{TCC3}
 R.~Brandenberger,
  ``Fundamental Physics, the Swampland of Effective Field Theory and Early Universe Cosmology,''
  arXiv:1911.06058 [hep-th];\\
  %%CITATION = ARXIV:1911.06058;%% 
  R.~Brandenberger,
  ``Trans-Planckian Censorship Conjecture and Early Universe Cosmology,''
  arXiv:2102.09641 [hep-th];\\
  %%CITATION = ARXIV:2102.09641;%%  
%\cite{Brandenberger:2021zib}
R.~Brandenberger,
``String Cosmology and the Breakdown of Local Effective Field Theory,''
[arXiv:2112.04082 [hep-th]];\\
%\cite{Brandenberger:2022pqo}
R.~Brandenberger and V.~Kamali,
``Unitarity Problems for an Effective Field Theory Description of Early Universe Cosmology,''
[arXiv:2203.11548 [hep-th]].   

\bibitem{BB}
%\cite{Brahma:2020zpk}
S.~Brahma, O.~Alaryani and R.~Brandenberger,
``Entanglement entropy of cosmological perturbations,''
Phys. Rev. D \textbf{102}, no.4, 043529 (2020)
doi:10.1103/PhysRevD.102.043529
[arXiv:2005.09688 [hep-th]].

\bibitem{noglobal}
%\cite{Banks:1989zw}
T.~Banks,
``Report on Progress in Wormhole Physics,''
Physicalia Mag. \textbf{12}, 19-68 (1990)
SCIPP-89-17.

\bibitem{swamprev}
T.~D.~Brennan, F.~Carta and C.~Vafa,
 ``The String Landscape, the Swampland, and the Missing Corner,''
 PoS TASI {\bf 2017}, 015 (2017)
 doi:10.22323/1.305.0015
 [arXiv:1711.00864 [hep-th]];\\
 %%CITATION = doi:10.22323/1.305.0015;%%
%\cite{Palti:2019pca}
E.~Palti,
``The Swampland: Introduction and Review,''
Fortsch. Phys. \textbf{67}, no.6, 1900037 (2019)
doi:10.1002/prop.201900037
[arXiv:1903.06239 [hep-th]];\\
 %\cite{vanBeest:2021lhn}
M.~van Beest, J.~Calder\'on-Infante, D.~Mirfendereski and I.~Valenzuela,
``Lectures on the Swampland Program in String Compactifications,''
Phys. Rept. \textbf{989}, 1-50 (2022)
doi:10.1016/j.physrep.2022.09.002
[arXiv:2102.01111 [hep-th]];\\
%\cite{Agmon:2022thq}
N.~B.~Agmon, A.~Bedroya, M.~J.~Kang and C.~Vafa,
``Lectures on the string landscape and the Swampland,''
[arXiv:2212.06187 [hep-th]].

\bibitem{noglobal-axion}
%\cite{Holman:1992us}
R.~Holman, S.~D.~H.~Hsu, T.~W.~Kephart, E.~W.~Kolb, R.~Watkins and L.~M.~Widrow,
``Solutions to the strong CP problem in a world with gravity,''
Phys. Lett. B \textbf{282}, 132-136 (1992)
doi:10.1016/0370-2693(92)90491-L
[arXiv:hep-ph/9203206 [hep-ph]];\\
%\cite{Kamionkowski:1992mf}
M.~Kamionkowski and J.~March-Russell,
``Planck scale physics and the Peccei-Quinn mechanism,''
Phys. Lett. B \textbf{282}, 137-141 (1992)
doi:10.1016/0370-2693(92)90492-M
[arXiv:hep-th/9202003 [hep-th]];\\
 %\cite{Barr:1992qq}
S.~M.~Barr and D.~Seckel,
``Planck scale corrections to axion models,''
Phys. Rev. D \textbf{46}, 539-549 (1992)
doi:10.1103/PhysRevD.46.539

\bibitem{Sarkar}
%\cite{Beyer:2022ywc}
K.~A.~Beyer and S.~Sarkar,
``Ruling out light axions: The writing is on the wall,''
SciPost Phys. \textbf{15}, no.1, 003 (2023)
doi:10.21468/SciPostPhys.15.1.003
[arXiv:2211.14635 [hep-ph]].

\bibitem{Dine}
%\cite{Dine:2022mjw}
M.~Dine,
``The Problem of Axion Quality: A Low Energy Effective Action Perspective,''
[arXiv:2207.01068 [hep-ph]].

\bibitem{RBfT}
%\cite{Brandenberger:1986jc}
R.~H.~Brandenberger,
``PHYSICS OF THE EARLY UNIVERSE: THE INFLATIONARY UNIVERSE AND COSMIC STRINGS,''
PRINT-87-0217 (CAMBRIDGE).

\end{thebibliography}
\end{document}